\documentclass[10pt, aps, pra, twocolumn, nofootinbib,superscriptaddress]{revtex4-1}
\usepackage{amsmath,amssymb,amsfonts}
\usepackage{algorithmic}
\usepackage{graphicx}
\usepackage{textcomp}
\usepackage{xcolor}
\usepackage{ragged2e}
\usepackage{booktabs, makecell, tabularx}
\usepackage{lipsum}

\usepackage{gensymb}
\usepackage{lineno}
\makeatletter
    \renewcommand\@make@capt@title[2]{%
     \@ifx@empty\float@link{\@firstofone}{\expandafter\href\expandafter{\float@link}}%
      {\textbf{#1}}\@caption@fignum@sep#2\quad}%
\makeatother
\makeatletter 
\renewcommand{\fnum@figure}{\textbf{Fig.~\thefigure}} 
\makeatother

\usepackage{xcolor}

\def\BibTeX{{\rm B\kern-.05em{\sc i\kern-.025em b}\kern-.08em
    T\kern-.1667em\lower.7ex\hbox{E}\kern-.125emX}}

\begin{document}

\author{Kaixuan Ye}
\thanks{These authors contributed equally to this work}
\affiliation{Nonlinear Nanophotonics Group, MESA+ Institute of Nanotechnology,\\
University of Twente, Enschede, Netherlands}
\author{Hanke Feng}
\thanks{These authors contributed equally to this work}
\affiliation{Department of Electrical Engineering \& State Key Laboratory of Terahertz
and Millimeter Waves, City University of Hong Kong, Hong Kong, China}
\author{Randy te Morsche}
\affiliation{Nonlinear Nanophotonics Group, MESA+ Institute of Nanotechnology,\\
University of Twente, Enschede, Netherlands}
\author{Akhileshwar Mishra}
\affiliation{Nonlinear Nanophotonics Group, MESA+ Institute of Nanotechnology,\\
University of Twente, Enschede, Netherlands}
\author{Yvan~Klaver}
\affiliation{Nonlinear Nanophotonics Group, MESA+ Institute of Nanotechnology,\\
University of Twente, Enschede, Netherlands}
\author{Chuangchuang Wei}
\affiliation{Nonlinear Nanophotonics Group, MESA+ Institute of Nanotechnology,\\
University of Twente, Enschede, Netherlands}
\author{Zheng Zheng}
\affiliation{Nonlinear Nanophotonics Group, MESA+ Institute of Nanotechnology,\\
University of Twente, Enschede, Netherlands}
\author{Akshay Keloth}
\affiliation{Nonlinear Nanophotonics Group, MESA+ Institute of Nanotechnology,\\
University of Twente, Enschede, Netherlands}
\author{Ahmet Tarık Işık}
\affiliation{Nonlinear Nanophotonics Group, MESA+ Institute of Nanotechnology,\\
University of Twente, Enschede, Netherlands}
\author{Zhaoxi Chen}
\affiliation{Department of Electrical Engineering \& State Key Laboratory of Terahertz
and Millimeter Waves, City University of Hong Kong, Hong Kong, China}
\author{Cheng Wang}
\email{cwang257@cityu.edu.hk}
\affiliation{Department of Electrical Engineering \& State Key Laboratory of Terahertz
and Millimeter Waves, City University of Hong Kong, Hong Kong, China}
\author{David Marpaung}
\email{david.marpaung@utwente.nl}
\affiliation{Nonlinear Nanophotonics Group, MESA+ Institute of Nanotechnology,\\
University of Twente, Enschede, Netherlands}

\date{\today}
\title{Brillouin photonics engine in the thin-film lithium niobate platform}

\begin{abstract}
Stimulated Brillouin scattering (SBS) is revolutionizing low-noise lasers and microwave photonic systems. However, despite extensive explorations of a low-loss and versatile integrated platform for Brillouin photonic circuits, current options fall short due to limited technological scalability or inadequate SBS gain. Here we introduce the thin-film lithium niobate (TFLN) platform as the go-to choice for integrated Brillouin photonics applications. We report the angle-dependent strong SBS gain in this platform, which can overcome the intrinsic propagation loss. Furthermore, we demonstrate the first stimulated Brillouin laser in TFLN with a tuning range $>20$~nm and utilize it to achieve high-purity RF signal generation with an intrinsic linewidth of 9 Hz. Finally, we devise a high-rejection Brillouin-based microwave photonic notch filter, for the first time, integrating an SBS spiral, an on-chip modulator, and a tunable ring all within the same platform. This TFLN-based Brillouin photonics engine uniquely combines the scalability of this platform and the versatility of SBS. Moreover, it bridges SBS with other functionalities in the TFLN platform, unlocking new possibilities for Brillouin-based applications with unparalleled performances.
\end{abstract}
\maketitle
\section*{Introduction}
Stimulated Brillouin scattering (SBS), arising from the coherent interaction between photons and phonons, is transforming integrated photonics \cite{Eggleton2019BrillouinPhotonics}. With its narrowband gain window, SBS is essential for high-selectivity filters \cite{Marpaung2015Low-powerSelectivity,Garrett2023IntegratedCircuit,Xu2024TunableBandwidth} and amplifiers \cite{Kittlaus2016LargeSilicon} in the next-generation optical and radio communications systems \cite{Marpaung2019IntegratedPhotonics}. Its unique acoustic dissipation mechanism enables sub-hertz linewidth integrated lasers \cite{Gundavarapu2019Sub-hertzLaser,Ko2024AResonators} and high-purity radio frequency (RF) signal generators \cite{Otterstrom2018ALaser,Klaver2024SurfaceChip}. Furthermore, manipulating its phase matching condition gives rise to on-chip non-reciprocal devices \cite{Kim2015non-transparency,Zhou2024NonreciprocalModes}.

Finding a versatile photonics platform that supports SBS is nevertheless challenging. Strong on-chip SBS gain requires large photoelastic coefficients of the material, simultaneous guidance of both the optical and acoustic waves, and low propagation loss of the platform. Although impressive proof-of-concept applications have been demonstrated, current platforms based on chalcogenide \cite{Pant2011On-chipScattering,Neijts2024On-chipWaves, Morrison2017CompactSilicon}, silicon-on-insulator \cite{Lei2024Anti-resonantChip, Kittlaus2016LargeSilicon, VanLaer2015InteractionNanowire, Dinter2024Anti-resonantInteraction, Munk2019SurfaceInsulator}, silicon nitride \cite{Gundavarapu2019Sub-hertzLaser,Gyger2020ObservationWaveguides,Botter2022Guided-acousticCircuitsb, Klaver2024SurfaceChip,Ji2024MultimodalityEffect}, and aluminum nitride \cite{Liu2019ElectromechanicalWaveguides} suffer from either insufficient SBS gain, material volatility, or structure instability. Consequently, these platforms fall short of the technological scalability in real-world applications.

Harnessing SBS in the thin-film lithium niobate (TFLN) platform can elevate the SBS technology to a new level of readiness. Recognized for its significant electro-optic (EO) coefficient, low loss, and good scalability \cite{Boes2023LithiumSpectrum,Zhu2021IntegratedNiobate}, novel applications including on-chip EO modulators \cite{Wang2018IntegratedVoltages}, optical frequency combs \cite{Zhang2019BroadbandResonator,Wu2024Visible-to-ultravioletWaveguides}, microwave photonic processors~\cite{Feng2024IntegratedEngine}, and integrated lasers~\cite{Snigirev2023UltrafastPhotonics,Guo2023UltrafastNiobate} with exceptional performances have all been demonstrated in this platform. While the optomechanical effect has been investigated in TFLN waveguides \cite{Sarabalis2021, Xie2024HighElectromechanics, Shao2019, Iyer2024CoherentDevices, Li2023FrequencyangularSteering}, investigation of SBS in the TFLN platform is still in its infancy \cite{Ye2023SurfaceWaveguides,Rodrigues2023On-ChipWaveguides,Yang2024}, and no Brillouin-based applications have been reported in this platform. A TFLN-based Brillouin photonics engine not only advances Brillouin-based applications onto a more scalable platform but also extends the capabilities of this platform, creating synergy with its existing functionalities.

\begin{figure*}[t!]
\centering
\includegraphics[width=\linewidth]{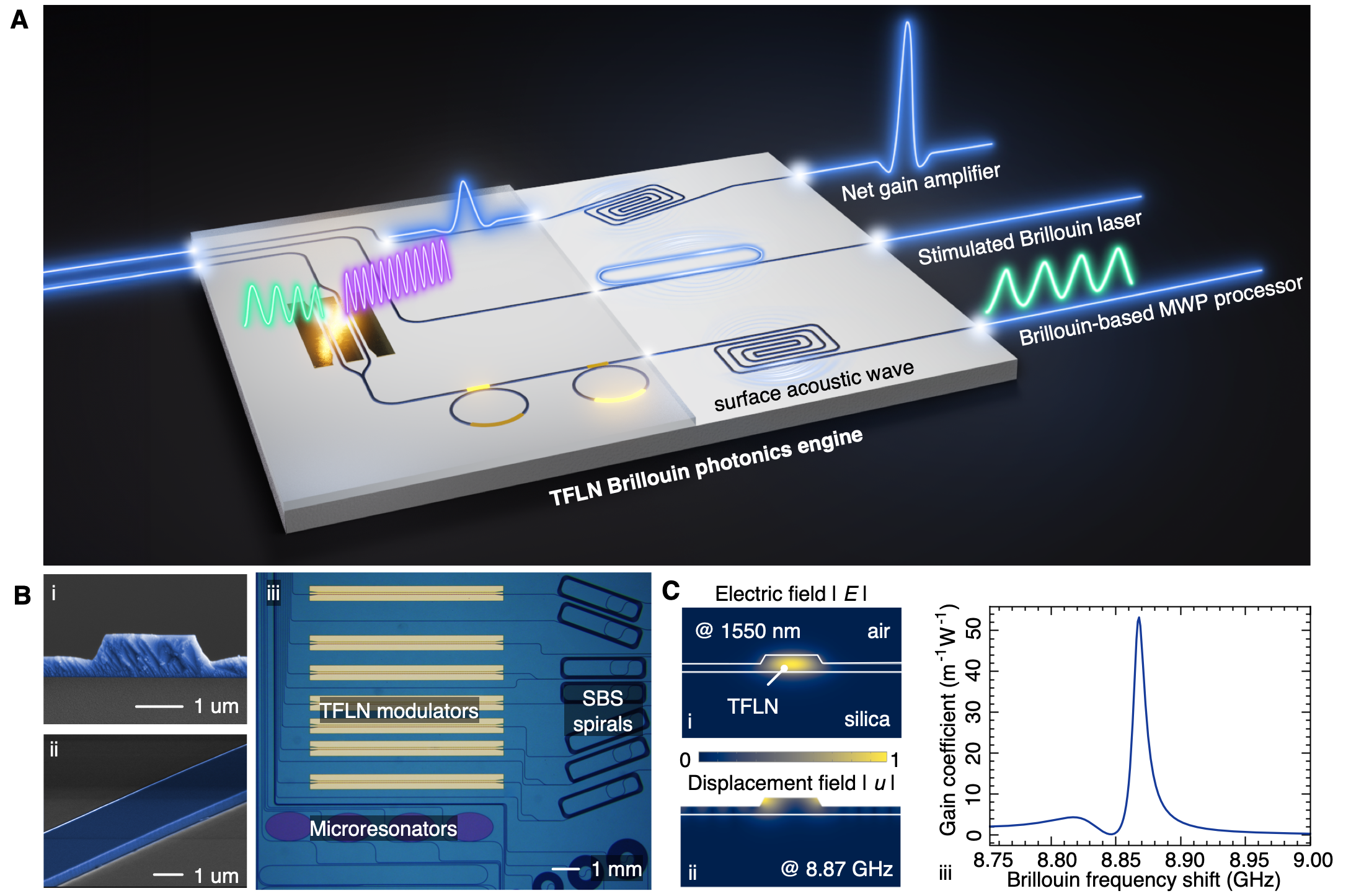}
\caption{\textbf{Brillouin photonics engine in the TFLN platform.} \textbf{(A)} Illustration of a TFLN circuit that contains high-speed electro-optic modulators, tunable rings and SBS-active waveguides, which can realize the net gain amplifier, the stimulated Brillouin laser, and the Brillouin-based microwave photonics processor. \textbf{(B)} \textbf{i.} Scanning electron microscope (SEM) image of the cross section of the half-etched TFLN waveguide. \textbf{ii.} SEM image of the sidewall, showing very small roughness. \textbf{iii.} Microscope image of a TFLN sample that integrates modulators, spirals, and resonators together. \textbf{(C)} Simulated SBS responses of the x-cut $0\degree$ TFLN waveguide, with the electric field, displacement field, and the gain coefficient profile shown in \textbf{i–iii}.}
\label{fig1}
\end{figure*}

Here we report the first observation and system applications of SBS in the TFLN platform. By leveraging the anisotropy of the material to achieve large photoelastic coefficients and to generate guided surface acoustic waves, we attain strong SBS gain in both x-cut and z-cut TFLN waveguides. This high SBS gain facilitates a versatile Brillouin photonics engine with multiple functionalities (Fig.\ref{fig1}A). First, we demonstrate a net internal gain amplifier that surpasses intrinsic propagation losses. Second, by incorporating the SBS gain in a high-quality racetrack resonator, we generate the first stimulated Brillouin laser (SBL) in TFLN. Lastly, we implement a high-rejection Brillouin-based notch filter with the modulator, the tunable ring, and the spiral all integrated in the TFLN platform for the first time.

\section*{Results}

\subsection*{Angle-dependent surface acoustic wave SBS}

Our investigation encompasses both z-cut and x-cut TFLN waveguides \cite{Ye2023SurfaceWaveguides}, which are commonly utilized across various applications. These waveguides are half-etched (Fig.\ref{fig1}B i. and ii.), striking a balance between tight optical mode confinement and low propagation loss. The standard waveguide structure facilitates the integration of high-speed TFLN modulators, centimeter-long spirals, and high quality-factor resonators on a single chip (Fig.\ref{fig1}B iii).

We attain strong SBS gain in these standard TFLN waveguides by leveraging the anisotropy of the lithium niobate. The Brillouin gain coefficient, which quantifies the strength of the SBS process, depends on photoelastic coefficients and the optoacoustic overlap. Recent theoretical work suggests the photoelastic coefficients of lithium niobate depend on the crystalline orientation of the wafer and the rotational angle of the waveguides \cite{Rodrigues2023StimulatedWaveguides}. Moreover, the mechanical properties are also angle-dependent, affecting the acoustic mode confinement. 

For uncladded TFLN waveguides at specific rotational angles, surface acoustic waves can be supported. Unlike bulk acoustic waves common in most SBS processes, surface acoustic waves exhibit a lower velocity, which enhances acoustic confinement and the optoacoustic overlap \cite{Klaver2024SurfaceChip, Neijts2024On-chipWaves, Zerbib2024StimulatedNanofibers}. This combination of high photoelastic coefficients and the surface acoustic wave results in a Brillouin gain coefficient exceeding 50~m$^{-1}$W$^{-1}$ (Fig.\ref{fig1}C).

\begin{figure*}[t!]
\centering
\includegraphics[width=\linewidth]{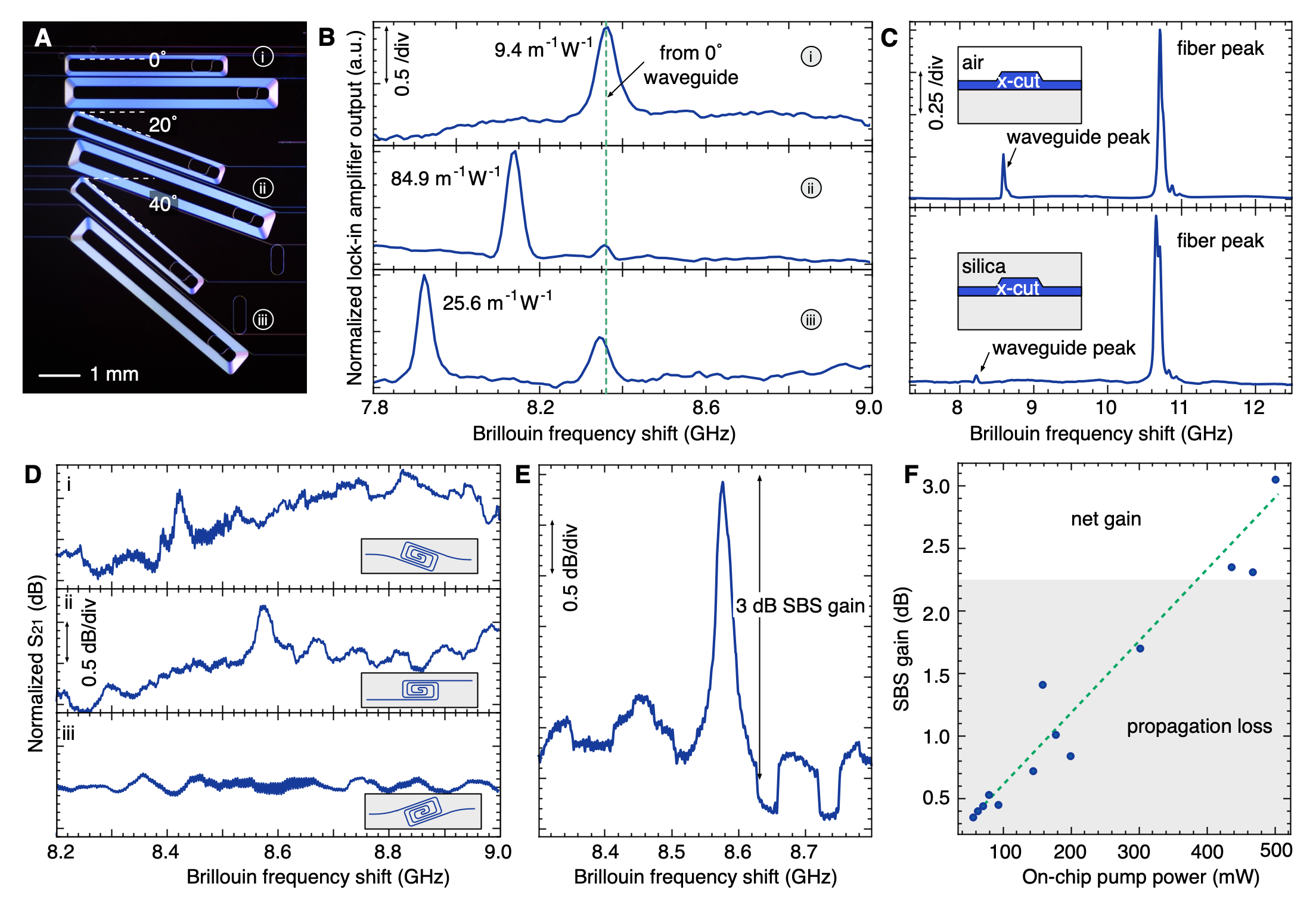}
\caption{\textbf{Angle-dependent surface acoustic wave SBS responses in the TFLN platform.} \textbf{(A)} Microscope image of a z-cut TFLN sample that contains waveguides with a rotational angle of $0\degree$, $20\degree$, and $40\degree$. \textbf{(B)} SBS responses of z-cut TFLN waveguides measured with a lock-in amplifier based setup. The peaks at 8.3~GHz in all three panels come from the $0\degree$ waveguide, while the other peak in \textbf{ii} and \textbf{iii} comes from the TFLN waveguide at different angles. Labels here match with the label in \textbf{(A)}, indicating the angle of the waveguide under test. \textbf{(C)} SBS responses of the uncladded and cladded x-cut TFLN waveguides. The stronger SBS response in the uncladded sample validates the role of the surface acoustic wave in enhancing the SBS responses in TFLN. \textbf{(D)} SBS responses of the x-cut TFLN waveguides measured with a VNA based setup.  VNA: vector-network analyzer. \textbf{(E)} SBS response of a 10~cm long  x-cut $0\degree$ TFLN waveguide, showing a SBS gain larger than 3~dB, overcoming the intrinsic propagation loss. \textbf{(F)} Measured SBS gain with increasing on-chip pump power. When the pump power is larger than 400~mW, the spiral under test shows net internal gain amplification.}
\label{fig2}
\end{figure*}

To experimentally validate the significant Brillouin gain coefficient in TFLN, we first characterized 1 cm long z-cut TFLN waveguides (Fig.\ref{fig2}A) using a lock-in amplifier based setup. Waveguides with varying rotational angles exhibit distinct SBS responses: the Brillouin frequency shift decreases from 8.36~GHz in the $0\degree$ waveguide to 8.14~GHz in the $20\degree$ waveguide and 7.92~GHz in the $40\degree$ waveguide. Notably, the Brillouin gain coefficient reaches 84.9~m$^{-1}$W$^{-1}$ in the $20\degree$ waveguide (Fig.\ref{fig2}B), underscoring TFLN as a promising platform for integrated Brillouin-based applications.

Furthermore, we verify the enhancement of the SBS response due to surface acoustic waves by comparing  x-cut TFLN waveguides with and without the top cladding. The uncladded waveguide exhibits a markedly stronger SBS response (Fig.\ref{fig2}C). Specifically, the Brillouin gain coefficient in the uncladded waveguide reaches 26.1~m$^{-1}$W$^{-1}$, nearly four times higher than the~7.0~m$^{-1}$W$^{-1}$ observed in the cladded waveguide, where surface acoustic wave is not supported.

We further characterized x-cut TFLN waveguides using a vector network analyzer (VNA) based setup, which can directly read out the on-off SBS gain. These x-cut waveguides also exhibit angle-dependent SBS responses, with the Brillouin gain coefficient increasing from 23.7~m$^{-1}$W$^{-1}$ at 8.44 GHz in the $20^\circ$ x-cut waveguide (Fig.\ref{fig2}D~i.) to 29.3 m$^{-1}$W$^{-1}$ at 8.57 GHz in the $0^\circ$ x-cut waveguide (Fig.\ref{fig2}D ii.). The Brillouin gain coefficient becomes too small to be extracted in the $-20^\circ$ x-cut waveguide (Fig.\ref{fig2}D iii.).

With increased pump power, we achieved net internal gain in a 10 cm long x-cut $0\degree$ spiral, surpassing the intrinsic propagation loss (Fig.\ref{fig2}E, F). However, the current fiber-to-waveguide coupling loss of 6.5~dB per facet limits the SBS gain from exceeding the total insertion loss. Optimized coupler designs \cite{Jia2023} could reduce this value to 1~dB per facet. Additionally, extending the length of the spiral and reducing the propagation loss of the waveguide could further boost the SBS gain, making it a unique narrowband amplifier in optical and radio communication systems.

\subsection*{Stimulated Brillouin laser demonstration}

\begin{figure*}[pt!]
\centering
\includegraphics[width=\linewidth]{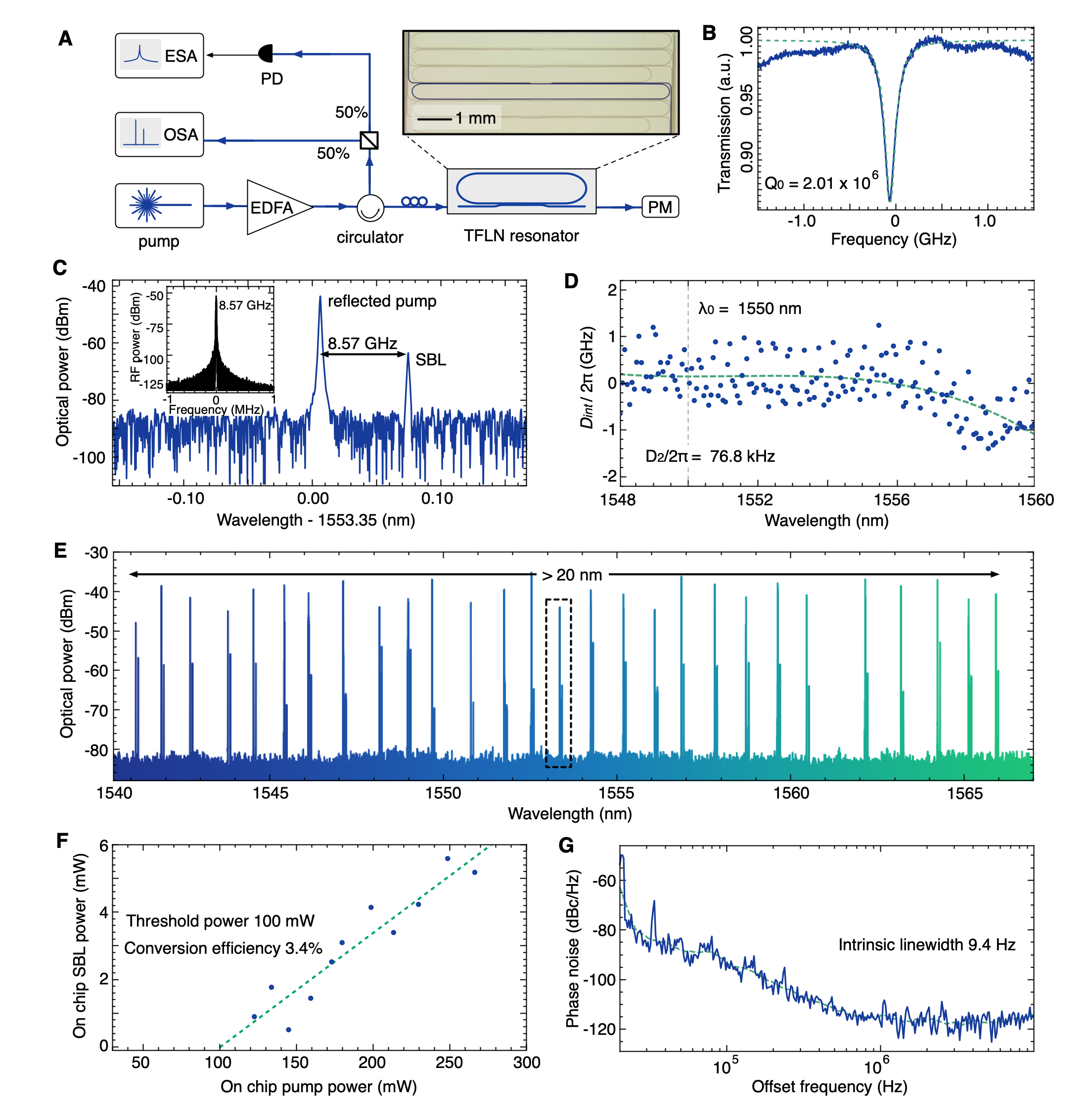}
\caption{\textbf{Stimulated Brillouin laser demonstration in a TFLN racetrack resonator.} \textbf{(A)} Experimental setup of the stimulated Brillouin laser demonstration. The pump laser is amplified with an EDFA before coupled into the TFLN resonator sample. The reflected light from the sample is monitored with the OSA, and also sent to the photodiode with an ESA for the RF spectrum measurement. EDFA: erbium-doped fiber amplifier; OSA: optical spectrum analyzer; ESA: electrical spectrum analyzer. \textbf{(B)} Measured optical resonance of the racetrack resonator. The intrinsic quality factor of this resonator is 2.01 million. \textbf{(C)}Optical spectrum of the reflected light at 1553.35~nm. The SBL is down shifted in frequency by 8.57~GHz. Inset: measured RF signal from the ESA. \textbf{(D)}~Measured integrated cavity dispersion $ D_{int}$ of the TFLN resonator. The low dispersion makes it possible for the wide tuning range of the SBL. \textbf{(E)} The SBL is tuned from 1540~nm to 1567~nm, with the tuning range only constrained by the operating range of the EDFA. The zoom-in plot of the spectrum within the dashed square is shown in \textbf{(C)}. \textbf{(F)}~Measured on-chip SBL power v.s. the on-chip pump power. The linear fitting gives a threshold power of 100~mW and a conversion efficiency from pump to SBL of 3.4~\%. \textbf{(G)}~ Measured single-sideband phase noise of the generated RF signal from the beating of the SBL and the pump. The extracted intrinsic linewidth of this signal is 9.4~Hz.}
\label{fig3}
\end{figure*}

\begin{figure*}[t!]
\centering
\includegraphics[width=\linewidth]{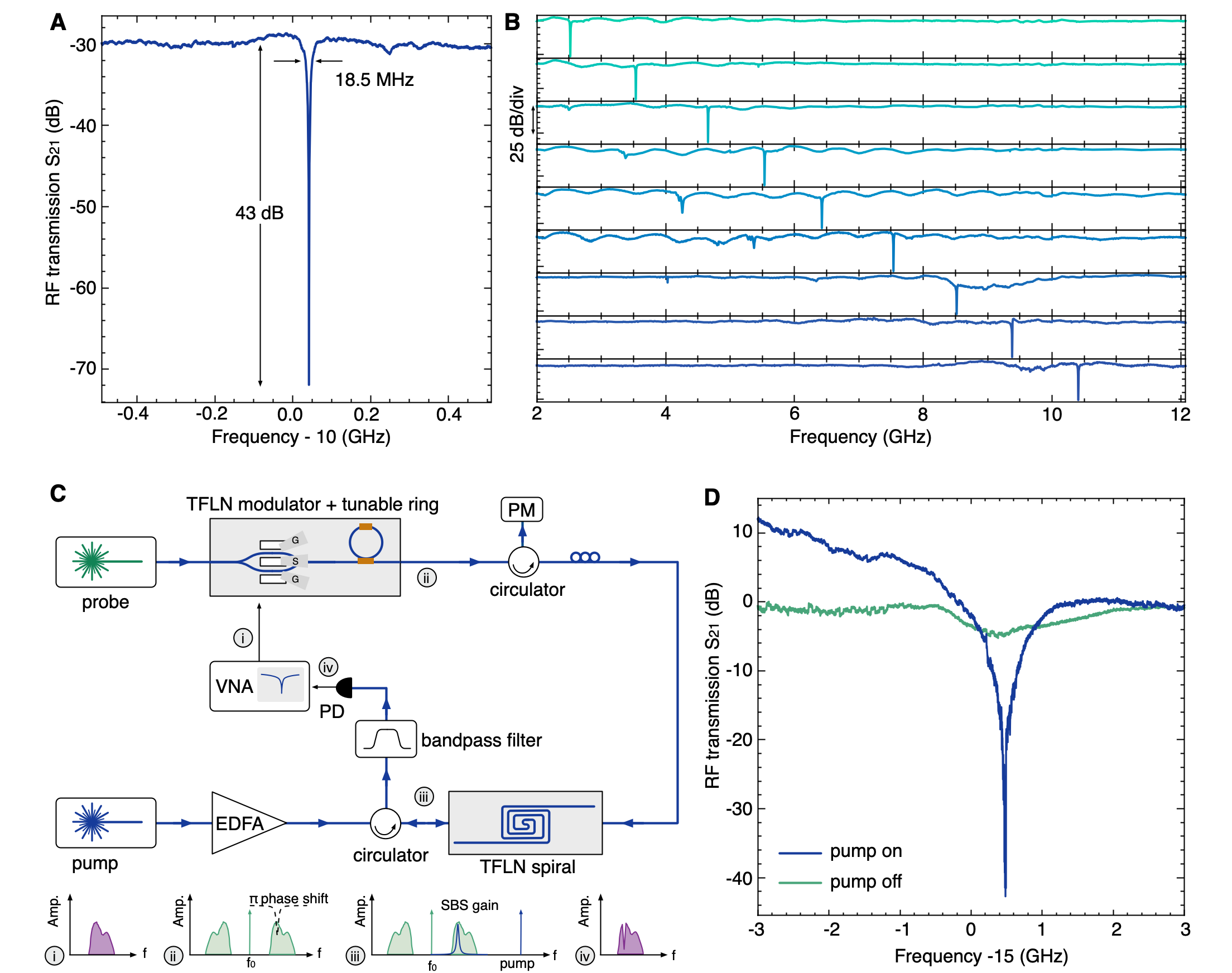}
\caption{\textbf{Brillouin-based microwave photonics filters demonstration.} \textbf{(A)} Measured high-rejection RF photonic notch filter response based on a 8 cm long x-cut spiral and an external in-phase quadrature (IQ) modulator. \textbf{(B)} The central frequency of this notch filter can be tuned over 8~GHz. \textbf{(C)} Experimental setup of a microwave photonic filter demonstration with an on-chip modulator, a tunable ring, and a spiral, all in the x-cut TFLN platform. Spectra at different points of the signal path: \textbf{i} The input RF signal applied to on-chip TFLN modulator. \textbf{ii} The probe is intensity modulated and its upper sideband is processed with an over-coupled ring resonator to apply $\pi$ phase shift at certain frequency. \textbf{iii} The SBS gain is applied to the upper sideband at the same frequency to compensate the attenuation induced by the ring. \textbf{iv} The resulting RF photonic notch filter. \textbf{(D)} Measured notch filter responses when the pump is turned on and off with the setup shown in \textbf{(C)}. }
\label{fig4}
\end{figure*}

Incorporating SBS gain within a high quality-factor resonator enables the first demonstration of the stimulated Brillouin laser (SBL) in the TFLN platform. The SBL is generated in a $0\degree$ x-cut TFLN racetrack resonator, with the setup illustrated in Fig.\ref{fig3}A. The intrinsic quality factor of the resonator reaches 2.01~million (Fig.\ref{fig3}B), and the free spectral range (FSR) of 8.58~GHz aligns precisely with the Brillouin frequency shift of the $0\degree$ x-cut TFLN waveguide. By tuning pump laser into the resonance of the racetrack resonator, we observed the counter-propagating SBL signal with the frequency down shifted by one FSR (Fig.\ref{fig3}C). 

The ultra-low dispersion of the TFLN waveguides facilitates a wide wavelength tuning range of the SBL. From the integrated cavity dispersion plotted in Fig.\ref{fig3}D, a group velocity dispersion ($D_2/2\pi$) as low as 76.8~kHz is calculated\cite{Ji2024}. This low dispersion ensures the FSR of the resonator closely matches with the Brillouin frequency shift across a wide wavelength range. Consequently, the SBL can be tuned from 1540~nm to 1567~nm (Fig.\ref{fig3}E), with the tuning range only constrained by the operational bandwidth of the erbium-doped fiber amplifier (EDFA) used in the experiment.

We further characterized the threshold power and the conversion efficiency of the SBL (Fig.\ref{fig3}F). We changed the on-chip pump power from 50~mW to 270~mW, and obtained an estimated threshold power of 100~mW and a conversion efficiency of 3.4\%. To further reduce the threshold power, the coupling coefficient from the bus waveguide to the resonator can be increased to near critical coupling. Moreover, a coupled ring molecule can be applied to reduce the mode volume, which also leads to a lower threshold power \cite{Ji2024MultimodalityEffect}.

The SBL in the TFLN platform holds great promise for high-purity RF signal generation. Unlike prior demonstrations in silicon nitride \cite{Gundavarapu2019Sub-hertzLaser}, chalcogenide \cite{Ko2024AResonators}, and bulk lithium niobate \cite{Nie2024Cross-polarizedPhotonics}, where the SBS linewidth is broader than optical cavity linewidth, the 20~MHz SBS linewidth in the $0\degree$ x-cut TFLN waveguide is narrower than the intrinsic linewidth of the racetrack resonator of 96~MHz. This unique regime of Brillouin lasing enables phonon linewidth narrowing, resulting in a pure RF signal generated from the beating note between the pump and the SBL. We measured the single-sideband phase noise of the generated RF signal with a self-heterodyne detection setup and obtained an intrinsic linewidth of 9.4~Hz (Fig.~\ref{fig3}G). This positions the SBL in the TFLN platform as a promising candidate for compact, low-noise RF oscillators. Additionally, technical noise at lower frequencies can be further reduced with laser stabilization techniques, such as using an optical reference cavity \cite{Sun2024IntegratedGeneration}, or a PDH lock circuit \cite{Liu2022Photonic-Resonators}.

\subsection*{Microwave photonic filters demonstration}
Leveraging the narrow linewidth of the SBS response in TFLN waveguides facilitates microwave photonic filters with high selectivity \cite{Marpaung2015Low-powerSelectivity}. Combining a 8 cm long $0\degree$ x-cut TFLN spiral with an external in-phase quadrature (IQ) modulator, we devised a notch filter with a linewidth as narrow as 18.5~MHz and the rejection larger than 43~dB (Fig.\ref{fig4}A). Moreover, by tuning the frequency of the pump laser, the Brillouin-based notch filter can be tuned from 2 to 10~GHz, which is limited by the bandwidth of the IQ modulator (Fig.\ref{fig4}B).

We advance this Brillouin-based microwave photonic notch filter technology by, for the first time, uniting an SBS spiral with a TFLN modulator and a tunable ring within the same platform. As shown in the experimental setup shown in Fig.\ref{fig4}C, we start by modulating the probe light with an TFLN intensity modulator. The upper sideband is then processed by a tunable, over-coupled ring on the same chip, which applies a $\pi$ phase shift at a designated frequency. This processed probe is subsequently coupled into a separate TFLN sample that contains a 8 cm long $0\degree$ x-cut TFLN spiral. The SBS gain from the spiral compensates the attenuation from the ring response while preserving the $\pi$ phase difference between the upper and lower sideband at the target frequency. Consequently, when the signal is sent to the photodiode, the resulting destructive interference produces a high-rejection notch filter (Fig.\ref{fig4}D). For further optimization and the spiral, high-speed modulator, and tunable rings has the potential to be all integrated on a single chip, leading to a higher link gain and improved noise figure.

\section*{Discussion}
In this work we harness strong SBS in the TFLN platform by exploiting the anisotropy of the material and the surface acoustic waves. The significant SBS gain makes the TFLN an ideal platform for developing a Brillouin photonics engine capable of diverse functionalities. As proof of concept demonstrations, we realized net-gain amplifiers, generated the first Brillouin laser, and devised an ultra-narrow bandwidth RF photonics notch filter, all within the standard x-cut TFLN platform. 

To further enhance performances, coupling loss between fiber and waveguide could potentially be reduced from 6.5~dB per facet to less than 1~dB per facet with dedicated spot-size converters \cite{Jia2023}. Additionally, anti-resonance waveguide structure could be applied to further enhance the SBS gain through improved acoustic confinement  \cite{Lei2024Anti-resonantChip}. Moreover, incorporating a coupled ring molecule design would allow for Brillouin laser generation with reduced threshold power, improved conversion efficiency, and a smaller device footprint.

Embedding this Brillouin photonics engine in a large-scale TFLN circuits could open the door to novel applications with unprecedented performances. For instance, the narrow SBS linewidth can be exploited to demonstrate a compact, high-resolution Brillouin optical spectrum analyzer \cite{Domingo2005}. In RF photonic systems, integrating the SBS spiral with high-speed TFLN modulators and tunable rings on a single chip could increase the link gain and lower the noise figure \cite{Wei2024programmable-niobate}. Furthermore, interaction between the SBS and Kerr effect in TFLN could lead to on-chip Brillouin-Kerr combs~\cite{Bai2021Brillouin-KerrMicroresonator}.

\section*{Methods}
\subsection*{Device fabrication}
The TFLN samples under test are fabricated with lithium niobate on insulator (LNOI) wafers from NANOLN. The lithium niobate film is 500~nm thick in the x-cut LNOI wafer, and 400~nm thick in the z-cut LNOI wafer. First, SiO$_2$ is deposited on the surface of a 4-inch LNOI wafer as an etching hard mask using plasma-enhanced chemical vapor deposition (PECVD). Waveguides are patterned with a UV stepper with a resolution of 500~nm. Afterwards, the patterns are transferred to the SiO$_2$ layer with a fluorine-based dry etching process, and then to the lithium niobate layer with an optimized Ar$^+$ based inductively coupled plasma (ICP) reactive-ion etching process. The lithium niobate is half-etched, leaving a 250~nm thick slab in x-cut wafer and 200~nm in the z-cut one. The wafer is annealed after the removal of the SiO$_2$ mask. 

\subsection*{SBS characterizations}
We characterized the SBS responses of the TFLN waveguides in two ways. First, a triple-intensity-modulated lock-in amplifier setup is applied. Details of this experimental setup can be found from our previous work \cite{Ye2024BrillouinInvited}. Second, we also implement a VNA-based setup. An single-sideband modulated probe is first coupled into the TFLN sample and the RF transmission is measured  as the baseline. Then a pump laser is coupled into the sample from the opposite direction, which induces SBS gain in the sideband of the probe. By comparing the two RF transmissions of the probe, the SBS gain from the TFLN waveguides can be directly read out.

\subsection*{Stimulated Brillouin laser demonstrations} 
An external cavity laser (Sacher TEC-520) is applied as the pump laser in the SBL experiments. The pump is amplified with an EDFA (Amonics AEDFA-37-R-FA) before coupled into the TFLN sample. The reflected light, including both the reflected pump and the generated SBL, is monitored with an optical spectrum analyzer (Finisar WaveAnalyzer 1550S). Part of the reflected light is also sent to a high-speed photodiode (Optilab PD 23-C-DC) and monitored with an electrical spectrum analyzer (Keysight N9000B).


\section*{Author Contribution}
D.M. and K.Y. developed the concept and proposed the physical system. K.Y., H.F., and R. M. designed the photonic circuits. H.F. and Z.C. fabricated the photonic circuits. K.Y., Y.K., and R. M. developed and performed numerical simulations. K.Y., A. M., R. M. and Cc. W. performed the experiments with input from Y. K., Z. Z., A. K., and A. T. I.. K.Y. and D.M. wrote the manuscript with input from everyone. D.M. and C.W. led and supervised the entire project.

\begin{acknowledgments}
The authors acknowledge funding from the European Research Council Consolidator Grant (101043229 TRIFFIC), Proof-of-Concept Grant (Veritas), Nederlandse Organisatie voor Wetenschappelijk Onderzoek (NWO) Start Up (740.018.021), the Research Grants
Council, University Grants Committee (N\_CityU113\/20, CityU11204022), and Croucher Foundation (9509005).
\end{acknowledgments}

\section*{Disclosures}
The authors declare no conflicts of interest.

\bibliographystyle{IEEEtran}
\bibliography{references}
\end{document}